\documentclass[journal=jpccck, manuscript=article,layout=twocolumn]{achemso}
\pdfoutput=1
\usepackage{graphicx}
\usepackage[english]{babel}
\usepackage{achemso}
\usepackage{amsmath}
\usepackage{hyperref}

		\author{D. Spirito}
		\affiliation{Nanochemistry department, Istituto Italiano di Tecnologia, Via Morego 30, I-16163 Genova, Italy}
		\alsoaffiliation{Graphene Labs, Istituto Italiano di Tecnologia, Via Morego 30, I-16163 Genova, Italy}
		\email{davide.spirito@iit.it}
		\author{S. Kudera}
		\affiliation{Nanochemistry department, Istituto Italiano di Tecnologia, Via Morego 30, I-16163 Genova, Italy}
		\author{V. Miseikis}
		\affiliation{Center for Nanotechnology Innovation @NEST, Istituto Italiano di Tecnologia, Piazza San Silvestro 12, I-56127 Pisa, Italy}
		\alsoaffiliation{Graphene Labs, Istituto Italiano di Tecnologia, Via Morego 30, I-16163 Genova, Italy}
		\author{C. Giansante}
		\affiliation{Center for Biomolecular Nanotechnologies @UNILE, Istituto Italiano di Tecnologia, Via Barsanti 1, 73010 Arnesano, LE, Italy}
		\alsoaffiliation{CNR NANOTEC-Istituto di Nanotecnologia, Polo di Nanotecnologia c/o Campus Ecotekne, via Monteroni, 73100 Lecce, Italy}
		\author{C. Coletti}
		\affiliation{Center for Nanotechnology Innovation @NEST, Istituto Italiano di Tecnologia, Piazza San Silvestro 12, I-56127 Pisa, Italy}
		\alsoaffiliation{Graphene Labs, Istituto Italiano di Tecnologia, Via Morego 30, I-16163 Genova, Italy}
		\author{R. Krahne}
		\affiliation{Nanochemistry department, Istituto Italiano di Tecnologia, Via Morego 30, I-16163 Genova, Italy}
		\alsoaffiliation{Graphene Labs, Istituto Italiano di Tecnologia, Via Morego 30, I-16163 Genova, Italy}

		\title{UV Light Detection from CdS Nanocrystal Sensitized Graphene Photodetectors at kHz Frequencies}

\begin{document}

\maketitle

\begin{abstract}
We have fabricated UV-sensitive photodetectors based on colloidal CdS nanocrystals and graphene. The nanocrystals act as a sensitizer layer that improves light harvesting leading to high responsivity of the detector. Despite the slow relaxation of the photogenerated charges in the nanocrystal film, faster processes allowed to detect pulses up to a repetition rate of 2 kHz. We have performed time-resolved analysis of the processes occurring in our hybrid system, and discuss possible photo-induced charge transfer mechanisms.\footnote{This document is the Accepted Manuscript version of a Published Work that appeared in final form in The Journal of Physical Chemistry C, 119, 42, 23859, copyright 2015 © American Chemical Society after peer review and technical editing by the publisher. To access the final edited and published work see \href{http://pubsdc3.acs.org/articlesonrequest/AOR-UT8wHWPSqtQ72WdMmjhm}{http://pubsdc3.acs.org/articlesonrequest/AOR-UT8wHWPSqtQ72WdMmjhm}}
\end{abstract}

\section{Introduction}
Graphene has established itself as a platform for electronics and optoelectronics because of its particular properties and versatility.\cite{dassarma_electronic_2011,avouris_graphene_2010,fiori_electronics_2014,koppens_photodetectors_2014} It has been exploited for photodetection, operating in a wide range of wavelengths. Here the main advantage of graphene is the intrinsically broad band and tunable absorption, which however yields only very low signal, and therefore limits the performance of pure graphene photodetectors \cite{koppens_photodetectors_2014}.

Recently, hybrid systems composed by graphene and a sensitizer layer have proved to be a promising solution to this limitation, for example with bulk semiconductors, \cite{an_tunable_2013,liu_quantum_2014} molecules, \cite{chen_biologically_2013} plasmonic structures, \cite{echtermeyer_strong_2011,yao_high_2014} and colloidal semiconductor nanocrystals (NC). \cite{klekachev_electron_2011,konstantatos_hybrid_2012,sun_infrared_2012}
In particular colloidal semiconductor nanocrystals are an attractive alternative due to their strong optical absorption that can be tuned via size and composition over a wide spectral range,\cite{talapin_prospects_2010} which in turn leads to high responsivity of the hybrid detectors.
Furthermore, the nanocrystal layer can be deposited by low cost  solution processes. \cite{konstantatos_hybrid_2012,sun_infrared_2012,li_photosensitive_2014} Here the working principle is that of a photoconductor: light is efficiently absorbed by the nanocrystal layer, generating electron-hole pairs. While carriers of one type (e.g., holes) are trapped in the nanocrystals, carriers of the other type (e.g., electrons) are transferred to graphene. The ratio of the relaxation time of the photogenerated charges to the transit time in the graphene channel can result in a large photoconductive gain. However, so far slow recombination processes limit the working speed of such devices below hundred Hz. \cite{koppens_photodetectors_2014}

Investigating the charge dynamics and the transfer mechanisms between the graphene and the nanocrystal layers is therefore important to improve the performance of the devices and to understand the fundamental limits. Previous works have focused on the effect of molecules absorbed onto nanocrystal surface, \cite{guo_oxygen_2013,zheng_visible_2013} and highlighted the role of ligands on the surface of nanocrystals.\cite{liu_graphene_2015}

In this work, we fabricated photodetectors based on colloidal CdS nanocrystal and CVD graphene from which we achieved high responsivity, fast detection in the kHz range and UV sensitivity. Pulsed laser illumination allows us to access the relevant time scales, and we propose a tentative description of the processes involved in the photoresponse.

\section{Experimental}

\subsection{Graphene growth and device fabrication}
The graphene was grown on 25~$\mu$m-thick Cu foil using low-pressure chemical vapor deposition (CVD). \cite{miseikis_rapid_2015} Initially, the foil was electropolished to remove surface impurities and reduce the surface roughness. It was then annealed for 10 minutes at 1060$^\circ$C and a pressure of 25~mBar in Ar atmosphere to increase the Cu grain size.
The growth of graphene was performed at 1060{$^\circ$C} by flowing 1 sccm of methane, 20 sccm of hydrogen and 980 sccm of argon for 10 minutes, producing isolated single-crystals of graphene of approximately 150 $\mu$m in diameter
The sample was then cooled to 120{$^\circ$C} before removing it from the reactor.

The graphene was transferred using the standard wet transfer technique. The foil was spin-coated with a 200 nm layer of poly(methyl-methacrylate) (PMMA) and immersed for 16 hours in a 0.1 M solution of $\mathrm{FeCl_3}$ to etch the copper. After the etching, a film of graphene and PMMA was left floating on the surface of the etchant solution.
The film was thoroughly rinsed by transferring it to a beaker of deionized water several times. The target substrate ($\mathrm{p-Si/SiO_2}$ chip, 285 nm oxide thickness) was then immersed in the water and the graphene/PMMA film was scooped out of the beaker. The substrate was left to dry under ambient conditions for 60 minutes.
Finally, the substrate was immersed in acetone for 4 hours to remove the PMMA support film and rinsed in isopropyl alcohol.

On these samples FETs have been fabricated using electron beam lithography; ohmic contacts (source and drain) were defined by evaporation of Cr/Au (5~nm/40~nm) and lift-off; devices were isolated by reactive ion etching in $\mathrm{Ar/O_2}$. Width and length of the channel of the FETs described in this work are $W=10~\mu$m and $L=20~\mu$m. The p-doped silicon substrate was contacted with indium to function as a planar back gate.

\subsection{Synthesis of CdS nanocrystals, ligand exchange and thin film deposition}
CdS nanocrystals with 3 nm diameter were synthesized following a protocol from Yu and Peng. \cite{yu_formation_2002} Briefly, 130 mg CdO, 63 ml octadecene and 4.35 ml oleic acid were mixed in a flask and the flask was connected to a Schlenk line. The mixture was heated to ca. 120$^\circ$C under vacuum, and kept at that temperature for 30 minutes.
Several purging cycles of nitrogen and vacuum were executed, and finally the flask was heated to 300$^\circ$C (under nitrogen) until the solution turned colorless. At that point the flask was cooled to 100$^\circ$C and vacuum was applied again in order to remove residual water that formed during the decomposition of CdO.
When no more fog was formed in the flask, the flask was filled with nitrogen again and the solution was heated to 293$^\circ$C.
At that point a solution of sulfur (24 mg) in degassed octadecene (15~ml) was injected into the flask. The temperature control was set to 250$^\circ$C after the injection. After 330 seconds, the growth of the nanocrystals was stopped by removing the heat source. Residual organic material was removed from the solution by repeated precipitation of the nanocrystals with methanol and redispersion in toluene.

For the deposition of the films the nanocrystals, the surface of the as-synthesized nanocrystals was chemically modified with p-methylbenzenethiolate ligands replacing native oleate ligands.\cite{giansante_darker_2015}. The nanocrystal were passivated with methylbenzenethiol and transferred into a mixture of chloroform and dichlorobenzene (2:3 volume ratio).\cite{giansante_darker_2015,giansante_colloidal_2013}
In details, this ligand exchange was performed by adding a methylbenzenethiol solution (95~mg methylbenzenethiol, 10~ml dichlorobenzene, 300~$\mu$l trioctylamine) in a 2:1 volume ratio to a 300~$\mu$M solution of CdS nanocrystals in toluene (concentration determined according to Ref. \cite{yu_experimental_2003}).
The success of the ligand exchange could be observed by a slight color change of the solution due to the decrease in exciton confinement caused by the methylbenzenethiol ligands.\cite{giansante_darker_2015} The nanocrystals were precipitated by the addition of hexane (ca. 1~ml per 100~$\mu$l of CdS solution) and redissolved in the solution of methylbenzenethiol. This process was repeated twice. Finally, the nanocrystals were dissolved in the mixture of chloroform and dichlorobenzene.

The thin films on the graphene devices were produced by covering first the substrate with the ligand-exchanged solution of CdS nanocrystals. The substrate with the solution was then covered with a small beaker for 10 minutes. After that, the beaker was removed and the substrate was spun at 300 RPM for 10 seconds and then at 1200 RPM and 2000 RPM for at least 15 minutes until the solvent had evaporated completely.

\begin{figure}
	\includegraphics[width=8.3cm]{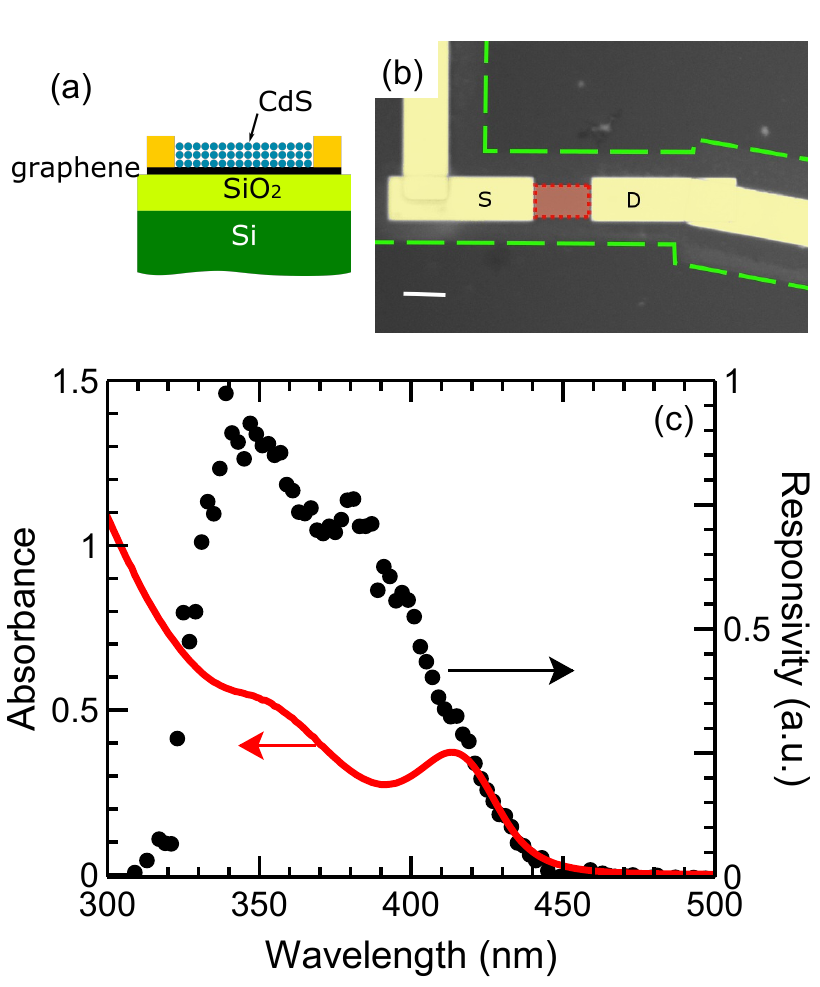}
	\caption{\label{fig:spectral}(a) Sketch of hybrid nanocrystals/graphene device (b) False color optical image of one of the investigated devices. The active area is marked by a dotted line and shaded in red. Ohmic contacts are marked by S (source) and D (drain). Green, dashed lines marks graphene isolated from the device. The scale bar is 15 $\mu$m. (c) Absorbance (left axis, red line) and responsivity (right axis, black dots) as a function of the wavelength for the CdS quantum dots/graphene hybrid device.}
\end{figure}

This procedure yields a 50 nm-thick sensitizer layer covering the conductive graphene channel.
A scheme of the final configuration of the device is reported in panel (a) of Fig. \ref{fig:spectral}; panel (b) is false color optical microscope image of one of the investigated devices.

Figure \ref{fig:spectral}c reports the absorbance of the nanocrystals recorded from toluene solution, and the spectral response of a NC coated graphene FET after the UV treatment that is described in detail in the following. The photocurrent spectra corresponds well to the optical absorption of the nanocrystals and confirms their key role in photo carrier generation.

\section{Results and discussion}

\begin{figure}
	\includegraphics[width=8.3cm]{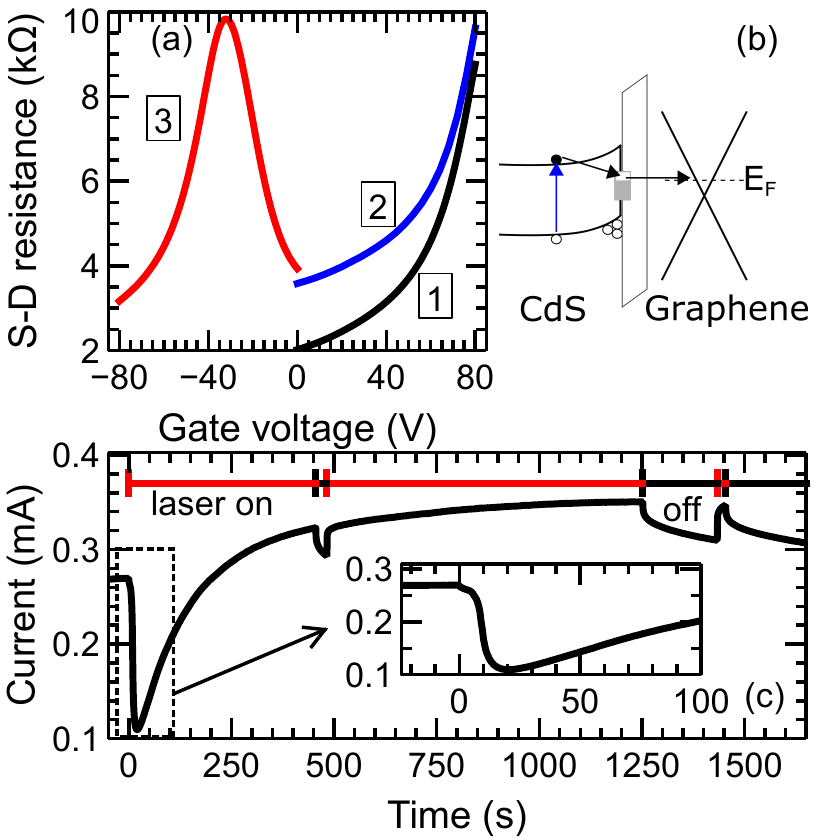}
	\caption{\label{fig:UVtreat}(a) Transfer characteristics of the graphene FET in the dark at the different steps of the experiments: 1. before deposition of nanocrystals; 2. after the deposition, before the UV treatment described in the text; 3. after the UV treatment. (b) Sketch of the bands and surface states in the CdS nanocrystal/graphene. (c) UV treatment procedure: source-drain current at $V_{SD}=1$~V and $V_{G}=0$~V, measured while a diode laser at 349 nm is turned on and off. The laser power was of 19.8~mW~cm\textsuperscript{-2}. The inset is a zoom of the initial part of the curve. }
	
\end{figure}

We measured the source-drain resistance as a function of the gate voltage at the different steps of the device fabrication and report the corresponding transfer characteristic in Fig. \ref{fig:UVtreat}a. All measurements were performed in vacuum.

After transfer and electrode fabrication the graphene is strongly p-doped (curve 1 in Fig. \ref{fig:UVtreat}a); similar p-doping was found after nanocrystal deposition (curve 2).
We then used a pulsed diode laser at 349 nm to illuminate the devices. Curve 3 in Fig. \ref{fig:UVtreat}a demonstrates that after illumination with the UV light (nanoseconds pulses at 1 kHz leading to an average intensity of 19.8~mW~cm\textsuperscript{-2}) the graphene changed to n-doped. We explore this process in more detail in Fig. \ref{fig:UVtreat}c, where we report the source-drain current measured during this UV-treatment.
At the beginning, the sample is in the dark. As the laser is turned on, the current decreases (negative photocurrent) and reaches a minimum value in about 20 s (see inset of Fig. \ref{fig:UVtreat}b), then the current increases and exceeds the value of the initial dark current, and eventually saturates after about 15 minutes. After passing through the minimum the photocurrent is positive, as shown at $t\sim 500~\mathrm{s}$ and $t\sim 1200~ \mathrm{s}$, and the Dirac peak has shifted to negative gate voltage (curve 3 in Fig. \ref{fig:UVtreat}a), which corresponds to n-doping. This n-doping is stable, and lasts as long as the sample is kept in vacuum.

The change from p- to n-doping and the related switching of the photocurrent polarity can be understood in terms of initial oxygen passivation of the NC surface. \cite{maserati_oxygen_2014,guo_oxygen_2013,zheng_visible_2013} At the beginning, the nanocrystals have oxygen adsorbed on their surface which can desorb under UV illumination, leaving positively charged surface traps behind, that we assume to be unpassivated $\mathrm{Cd^{2+}}$ atoms. During this process, photogenerated electrons will screen the positive surface traps and at the same time the traps can mediate electron transport into the graphene (see a sketch of the process in Fig. \ref{fig:UVtreat}b).
This process raises the Fermi level in graphene, approaching the Dirac point and thus first reducing the hole conductivity, and after crossing the Dirac point (minimum in Fig. \ref{fig:UVtreat}c) conduction is mediated by electrons and increases with illumination until the electron transfer saturates. When light is turned off under this condition, the electron density in graphene decreases, photocurrent is positive, and the surface traps on the NC surface cannot efficiently be passivated under vacuum. Thus, the graphene layer does not return to its original p-doped status when the light is turned off, but remains n-doped. Exposing the sample to air restores the original p-doping.

\subsection{Photoresponse measurements}
\begin{figure}
	\includegraphics[width=8.3cm]{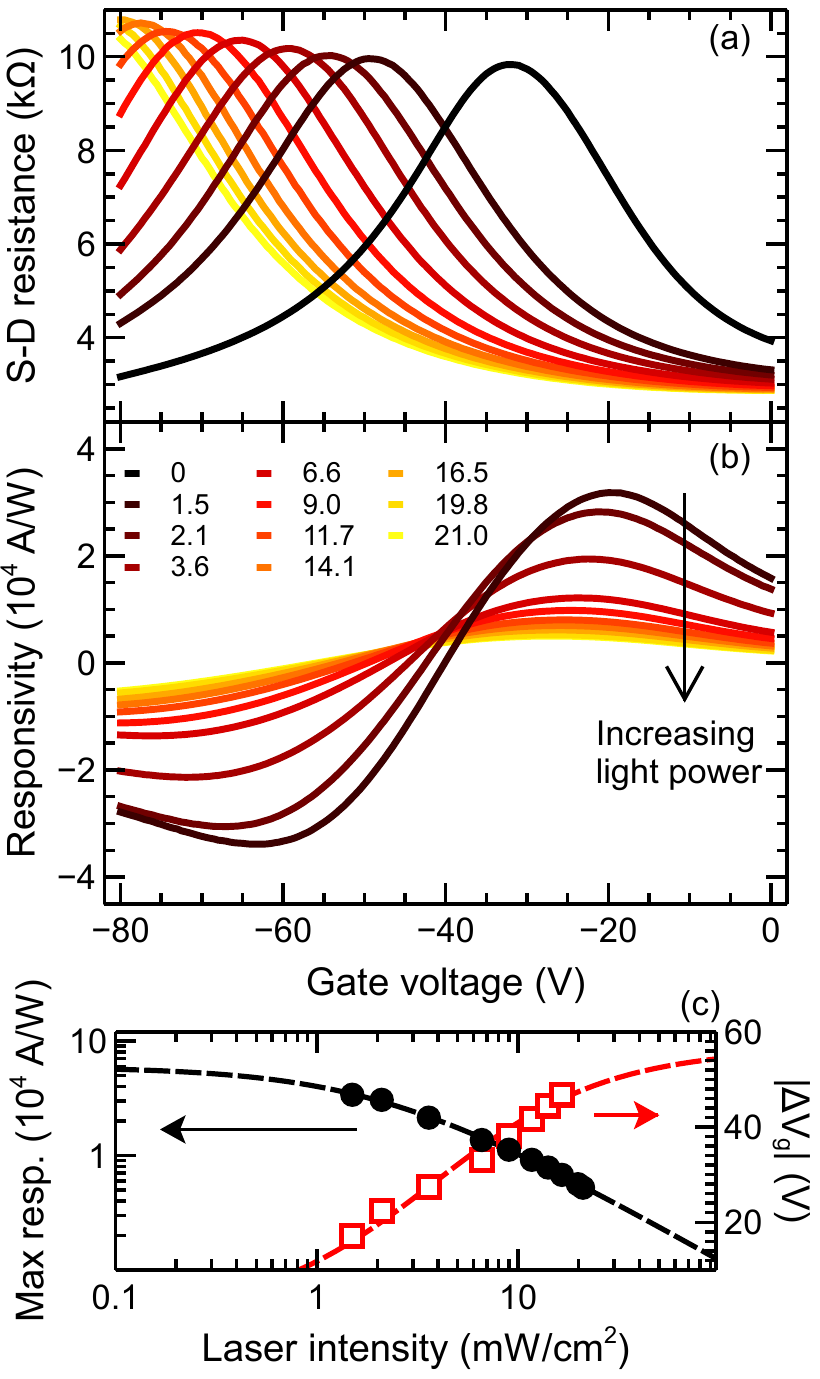}
	\caption{\label{fig:responsivity}(a) Source-drain resistance versus gate voltage at different laser intensity (349 nm wavelength), recorded at $V_{SD}=1$ V. The Dirac peak shifts toward more negative voltages with respect to the dark curve as the light power increases (see legend in panel b). 
		(b) Responsivity at 349 nm as a function of gate voltage. The legend states the laser intensity in mW cm\textsuperscript{-2} for panels (a,b).
		(c) Maximum responsivity (full circles, left axis) and gate voltage shift (empty squares, right axis) as a function of laser intensity. Dashed lines are fits to the equation \ref{eq:respP} and \ref{eq:DvgP} in the text.
	}
\end{figure}

The source-drain resistance as a function of the gate voltage is reported in Fig. \ref{fig:responsivity}a, acquired in dark and with increasing laser intensity. Upon UV-treatment, the graphene in the dark is n-doped; the mobility and the residual carrier density are $\mu=1100$ cm\textsuperscript{2} V\textsuperscript{-1} s\textsuperscript{-1} and $n_0=10^{12}$ cm\textsuperscript{-2}, respectively, calculated using the formula in Ref. \cite{kim_realization_2009}.  As the laser power increases, the peak (Dirac point) shifts toward more negative voltage, indicating that electrons are transferred from the nanocrystals. $\mu$ and $n_0$ do not change (within 10\%) with illumination.

Fig. \ref{fig:responsivity}b shows the responsivity, defined as $\mathcal{R}=(I_{light}-I_{dark})/P$ ($P$ being the power impinging on the device), as a function of the gate voltage. As the gate voltage is swept toward negative values, the photocurrent (hence the responsivity) is positive and reaches a maximum. Then it goes to zero and changes sign reaching a minimum (negative) value and then again approaches zero. 

We can model the source drain resistance versus gate voltage by: \cite{kim_realization_2009}
\begin{equation}
	\label{eq:kim}
	R\left(V_G\right)=R_{c} + \frac{1}{\alpha \sqrt{n_{0}^{2} + n_{V}^{2}}},
\end{equation}

where $R_c$ is the contact resistance, $n_V=(C_G/e)\left(V_G-V_D\right)$ is the carrier density induced by the gate voltage ($C_G$ being the gate capacitance per unit area, $V_D$ the Dirac voltage), $n_0$ the residual carrier density, and $\alpha=e\mu W/L$ ($\mu$ being the mobility, and $W$, $L$ being width and length of the FET channel).

The effect of the light is to transfer charge to the graphene. Therefore, we model it replacing $n_V\rightarrow n_V+\Delta n$, or $V_D\rightarrow V_D-(e/C_G)\Delta n$.

With these definitions, the photocurrent reads
	\begin{align}
	I_{ph}=I_{light}-I_{dark}=&\frac{V_{SD}}{R_{c} + \frac{1}{\alpha \sqrt{n_{0}^{2} + \left(\Delta n + n_{V}\right)^{2}}}} -\nonumber\\ &- \frac{V_{SD}}{R_{c} + \frac{1}{\alpha \sqrt{n_{0}^{2} + n_{V}^{2}}}}.
	\end{align}

The zero of the photocurrent occurs when $n_V=-\Delta n/2$, independently of the other parameters. We experimentally observe this dependence on the photogenerated carrier density as the shift in the zero of the responsivity.
We can also account for our observation that the photocurrent (and thus the responsivity) approaches zero far away from the Dirac point (large $n_V$). In this case, the responsivity is limited by the contact resistance $R_c$, and one finds
\begin{equation}
	\label{eq:iph_rc0}
	I_{ph}\approx\frac{\Delta n}{\alpha n_{V}^{2} R_{c}^2} V_{SD}\rightarrow 0,\quad n_V\rightarrow\infty,
\end{equation} 
while, if $R_c=0$, the photocurrent approaches a constant value
\begin{equation}
	I_{ph}\approx \frac{W}{L}e\mu\Delta n V_{SD}\neq 0, \quad n_V\rightarrow\infty.
\end{equation}
Figure \ref{fig:responsivity}b shows a clear impact of Rc since far away from the maximum it approaches zero. However, in order to find a simple expression to fit the maximum of the responsivity in Fig. \ref{fig:responsivity}b versus light power, we take $R_c=0$, which is justified since it plays only a role far away from the maximum.
Furthermore, we need to elaborate on the light power dependence of the photoexcited charges ($\Delta n$) in the graphene channel.

In general, the number of charge excited in the nanocrystals per unit area is
\begin{equation}
\label{eq:nnc}
	n_{NC}=\frac{\eta}{WL} \frac{P}{h \nu}\tau,
\end{equation}

where $\eta$ is excitation efficiency (including the absorption coefficient), $P$ is the power impinging on the device, $\tau$ the lifetime of excited carrier.
The charges induced in  the graphene layer are given by $\Delta n=k n_{NC}$, where $k$ is introduced to account for non-perfect efficiency in the process.

In Fig. \ref{fig:responsivity}c we report the dependence of the maximum responsivity and the shift of the Dirac point (that is proportional to $\Delta n$) versus light intensity. In the simplest case the photogenerated charges are directly proportional to the light power.
However, in our data we find non-linear behavior, indeed, we observe that the responsivity decreases as the power increases, while with a linear dependence one would expect a constant responsivity following the photocurrent dependence on $P$: $I_{ph}\propto \Delta n\propto P$, $\mathcal{R}=I_{ph}/P=const$.

Another approach used by other groups is a simple power law ($\Delta n\propto P^\beta$),\cite{sun_infrared_2012,wang_wide_2015} but also this does not reproduce our data. Ref. \cite{guo_oxygen_2013} proposes to explain the saturating behavior of the responsivity based on a power-dependent lifetime of the photogenerated charges, given by
\begin{equation}
	\label{eq:tauP}
	\tau\left(P\right)=\tau_0\frac{1}{1+\left(P/P_0\right)^\beta}.
\end{equation}

For low power, the lifetime saturates to a finite value $\tau_0$.  Here $\beta$ describes the dependence at high power. A similar formula was considered also in Ref. \cite{konstantatos_hybrid_2012}, with $\beta=1$, which worked also for experimental data fits in Ref. \cite{guo_oxygen_2013}. In our case, we can fit the power dependence of the responsivity in Fig. \ref{fig:responsivity}c 
using Eq. \ref{eq:iph_rc0}, \ref{eq:nnc} and \ref{eq:tauP} with $\beta=1$, with the formula

\begin{equation}
	\label{eq:respP}
	\mathcal{R} = \frac{I_{ph}}{P} = \frac{1}{P}\frac{W}{L}e\mu\Delta n V_{SD}=\frac{e}{h \nu}\frac{k\eta\tau_0}{\tau_{tr}}\frac{1}{1+P/P_0},
\end{equation}
where $\tau_{tr}=L^2V_{SD}^{-1}\mu^{-1}$ is the transit time of the carriers in the FET channel. The Dirac point shift is given by
\begin{equation}
	\label{eq:DvgP}
	\Delta V_G = \frac{e}{C_G}\Delta n = \frac{e}{C_G WL}\frac{1}{h \nu} \left(k\eta\tau_0\right) \frac{P}{1+P/P_0}.
\end{equation}

We note that this model was proposed to describe the saturation of trap states on nanocrystal surfaces by photoexcited carriers, which is in line with the processes that we assume relevant for our system.

As for the figures of merit of these devices, the responsivity has a peak value of $3.4\cdot10^{4}$ A/W at 1.5 mW/cm\textsuperscript{2}, and from this value the external quantum efficiency can be evaluated as $EQE=\mathcal{R}h\nu e^{-1}$, reaching $10^5$. UV sensitivity of our device is demonstrated by a 25-times smaller responsivity in the visible range (measured at 473 nm).
With respect to previously reported works, we note that  devices of similar concept, based on simple deposition of colloidal nanocrystals, have reached $10^{4}$ A W$^{-1}$ in the near UV-blue range,\cite{huang_photoelectrical_2013,guo_oxygen_2013,son_photoresponse_2013} with a time response that was of the order of seconds or fractions of seconds, while higher responsivity values can be reached with subsequent chemical process or more complex substrate structures.\cite{dang_ultrahigh_2015,shao_organic_2015}

The noise equivalent power (NEP) is given by the current noise divided by the responsivity. The noise can be evaluated experimentally from a time trace of the dark current. This yields 10\textsuperscript{-12}~W~ Hz\textsuperscript{-1/2} as minimum NEP at the maximum responsivity, which corresponds to a detectivity (defined as $\sqrt{S}/NEP$, where $S$ is the surface of the device) of 10\textsuperscript{9}~Jones.
We note, however, that this estimation includes both the intrinsic noise of the device and the noise in the experimental set-up, which was not optimized for such purposes. If we assume only thermal noise, we can estimate the noise current as $\sqrt{4k_BT/R_d}$, $R_d$ being resistance in the dark, $T$ the temperature. With this estimate, NEP and detectivity  improve to 10\textsuperscript{-17}~W~Hz\textsuperscript{-1/2} and 10\textsuperscript{13}~Jones, respectively.

\begin{figure}
	\includegraphics[width=8.3cm]{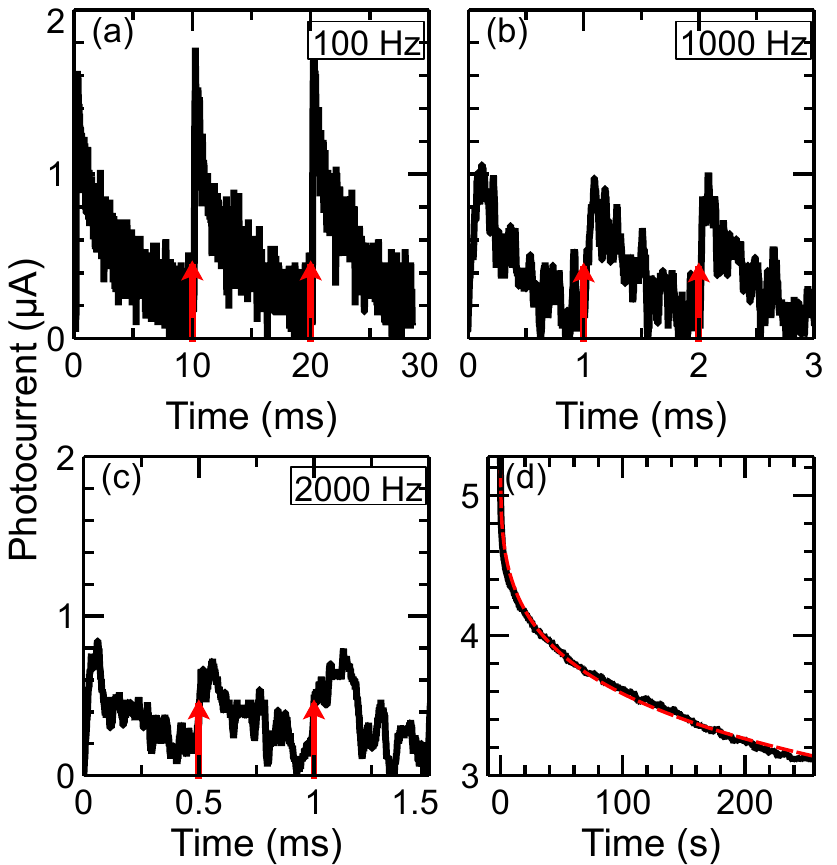}
	\caption{\label{fig:pulses_panel}Detection at $V_G=0~\mathrm{V}$ and $V_{SD}=1~\mathrm{V}$ of short laser pulses at repetition rates (a) 100 Hz, (b) 1000 Hz, (c) 2000 Hz. Red arrows mark the nanosecond pulses. The pulse energy is 18.6, 15.9, 13.8 $\mathrm{\mu J\,cm^{-2}}$ for 100, 1000, 2000 Hz. (d) Decay current after turning off the laser on a longer time range. The red dashed curve is a fit with the stretched exponential model described in the text.}
\end{figure}

To assess the temporal response of our photodetectors experimentally, we have performed pulsed laser excitation experiments at different frequencies and with varying fluences. Pulse width was some nanoseconds.
Figure \ref{fig:pulses_panel}a-c shows time traces of the photocurrent for pulsed stimulation at different frequencies, which clearly demonstrate that the device can detect the laser signal up to 2 kHz (Fig. \ref{fig:pulses_panel}c), which is at least one order of magnitude faster than the values reported for sensitized graphene photodetectors, \cite{konstantatos_hybrid_2012,guo_oxygen_2013,liu_graphene_2015} and several orders of magnitude faster than NC based photodetectors.\cite{konstantatos_pbs_2007}
From the fitting of the data in Fig. \ref{fig:responsivity}c we can extract a response time of our system in the form of the product $k\eta\tau_0$, yielding $k\eta\tau_0\sim 0.8$~ms for the responsivity fit, and $k\eta\tau_0\sim 0.7$~ms for the fit of the Dirac point shift. 
At least for the fastest response of our detectors these values appear to be in the correct range. We note that $k\eta$ will be smaller than 1, thus our fitting  represents a lower limit for the charge carrier relaxation time $\tau_0$.

Figure \ref{fig:pulses_panel}d reports the time trace of the photocurrent after illumination is turned off on a longer time period. 
In this case, we observe a much longer timescale for the decay, which we can describe as a persistent photocurrent effect.
A simple exponential decay does not reproduce the experimental curve, since the current decay is characterized by an initial fast decrease, followed by a much slower tail. Although a sum of two exponential functions would improve the fit, we can as well use the stretched exponential function $I(t)\propto\exp\left(-(t/\tau_s)^\beta\right)$, often used to describe the time response in detectors based on nanocrystals and other materials.\cite{mondal_observation_2011,carrey_photoconductivity_2008,ginger_charge_2000}
A system described by this formula can be interpreted as a continuous distribution of time scales which sum up in the response.\cite{johnston_stretched_2006} Within this model, we obtain the best fit with an average timescale of about 2000 s.

\begin{figure}
	\includegraphics[width=8.3cm]{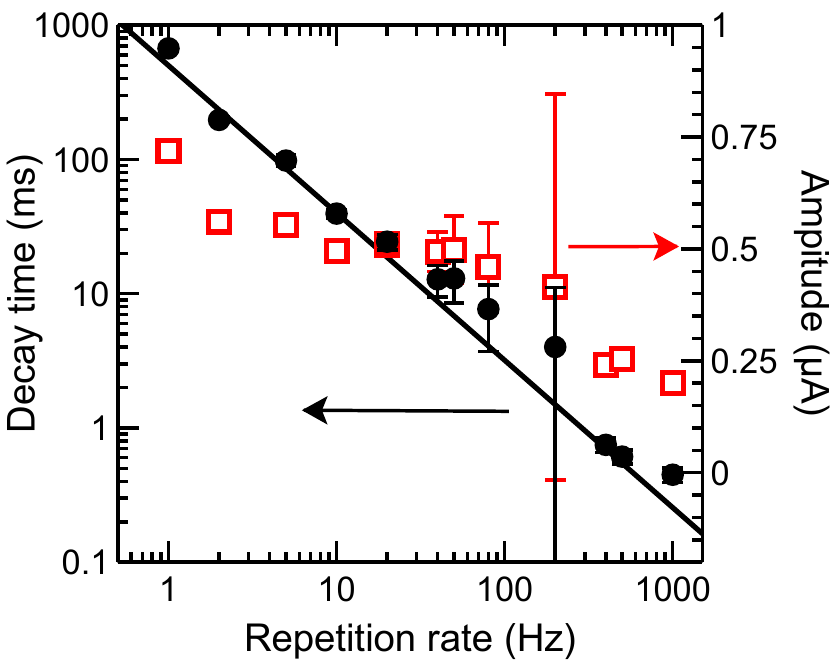}
	\caption{\label{fig:decayRate}Decay time (full circles, left axis) and amplitude (empty squares, right axis) of the photocurrent response at different repetition rates for laser pulses. The device was at $V_{SD}=1$~V and $V_G=0$~V. The energy of the pulses is 4.8~ $\mu$J~cm\textsuperscript{-2}. The black line is a fit of the decay time with $\tau\propto \mathrm{rate}^{-a}$, resulting in $a=1$.}
	
\end{figure}

To gain more insight into the time constants that play a role in our system we analyzed the photocurrent time traces at different repetition rate while keeping pulse power, $V_G$ and $V_{SD}$ constant.
The time traces were fitted with a single exponential adjusting the amplitude, the decay time and the base current, and the results are reported in Fig. \ref{fig:decayRate}. We note that single exponential fits worked well at high frequencies, while at lower frequency ($<$5~Hz) they only described the fastest component of the trace.
The good single exponential fits at high frequency are possible because the slower processes contribute only as a DC offset (a persistent photocurrent).
Figure \ref{fig:decayRate} shows that the decay time is inversely proportional to the repetition rate and suggests a continuous distribution of time scale $\tau$. Consistently, the amplitude does not show the typical $1/f$ relation expected for a simple single relaxation time system. 

It is well known that the optoelectronic properties of colloidal nanocrystals depend significantly on their surface chemistry.\cite{talapin_prospects_2010} While ligands coming from the synthetic procedures result in good surface passivation and colloidal stability in solution, they are detrimental for charge transport.
We exchanged these native ligands and worked with methylbenzenethiolate-capped nanocrystals in our devices that yield thin-films with enhanced conductivity.\cite{giansante_colloidal_2013}

With such short ligands the nanocrystal surface becomes more accessible for the photogenerated charges, and shallow surface traps will play a strong role concerning recombination processes. On the other hand, the UV assisted desorption of oxygen molecules under vacuum is likely to result in unpassivated  $\mathrm{Cd^{2+}}$ atoms on the NC surface, which should lead to deep trap states approximately 0.7 eV below the conduction band.\cite{chrysochoos_recombination_1992,lu_trap_2009,zhao_photoluminescence_2012}

Konstantatos et al.\cite{konstantatos_pbs_2007} have reported that trap states significantly influence the relaxation times of photogenerated carriers in NC films. They demonstrated that deep traps around 0.3 eV below the conduction band lead to long relaxation times of few seconds, while shallow traps of less than 0.1 eV resulted in short times responses of few tens of milliseconds.
Consequently, trap state engineering could be exploited as a means to reduce the response time of PbS NC photodetectors to 25 ms by using ethanethiols as surface ligands.\cite{konstantatos_engineering_2008}
We therefore tentatively assign the large distribution of relaxation times observed from our hybrid graphene-NC photodetectors to surface trap states with different depth. Here the shallow trap states enable the extremely fast response times of less than a millisecond, while deep traps are responsible for the observed persistent photoconductivity. 

\section{Conclusions}
We presented semiconductor nanocrystal sensitized graphene photodetectors that can provide high sensitivity up to $10^4~\mathrm{A/W}$ and fast signal detection reaching the kHz range. 
Here the overcoating of graphene FETs with a layer of CdS nanocrystals by simple spin coating greatly enhanced their absorption in the UV spectral range.
The surface passivation of the nanocrystals with methylbenzenethiolate ligands resulted in a photocurrent decay spanning timescales from hundreds of microseconds to minutes.

The fast decay times enabled detection at high repetition rates well above the limit needed, for example, for video recording. This approach can be extended straightforwardly to constitute a platform of semiconductor nanocrystals passivated with arenethiolate ligands, such as Cd-, Pb-, and Zn-based chalcogenides, that enables tuning of the spectral sensitivity of such hybrid photodetectors via the nanocrystal material, while maintaining high sensitivity and fast response.
Future developments will aim at increasing the responsivity at high repetition rate by eliminating deep surface traps, and by the use of less toxic nanocrystal materials based on copper indium (gallium) selenides, for instance.

\begin{acknowledgement}
This work was supported by the EC under the Graphene Flagship program (contract no. CNECT-ICT-604391).
\end{acknowledgement}

\bibliography{refs} 
\bibliographystyle{achemso}

\end{document}